\documentclass[a4paper,amsmath,amssymb]{jpconf}
\usepackage{graphicx}
\usepackage{amsmath, amsthm, amssymb}
\usepackage{color}

\newcommand{\id}{\textrm{d}}

\newcommand{\bbZ}{{\mathbb Z}}

\def\bea{\begin{eqnarray}}
\def\eea{\end{eqnarray}}
\def\ba{\begin{array}}
\def\ea{\end{array}}
\def\n{\nonumber}

\def\la{\langle}
\def\ra{\rangle}

\begin{document}
%\title{Frenetic Aspects of Response}
\title{Nonequilibrium Response and Frenesy}

\author{Urna Basu and Christian Maes}

\address{Institute for Theoretical Physics, KU Leuven}

\ead{urna.basu@fys.kuleuven.be}

%%%%%%%%%%%%%%%%%%%%%%%%%%%%%%%%%%%%%%%%%%%%%%%%%%%%%%%
%%                                                                              Abstract                                                                                             %%
%%%%%%%%%%%%%%%%%%%%%%%%%%%%%%%%%%%%%%%%%%%%%%%%%%%%%%%
\begin{abstract}
We present examples of how time-symmetric kinetic factors contribute to the response either in nonlinear order around equilibrium or in linear order around nonequilibrium.  The phenomenology we associate to that so called frenetic contribution are negative differential conductivity, changes in the Einstein relation between friction and noise, and population inversion.
\end{abstract}
%%%%%%%%%%%%%%%%%%%%%%%%%%%%%%%%%%%%%%%%%%%%%%%%%%%%%%%%%
\section{Introduction}
\label{sec:Introduction}
%%%%%%%%%%%%%%%%%%%%%%%%%%%%%%%%%%%%%%%%%%%%%%%%%%%%%%%%%
Understanding how a system responds to an external stimulus is both interesting and important. Questions include how to predict it and to identify on what it depends. We could for example suspect from general intuition that the response of an observable would be related to its variance; the system is more susceptible to external stimuli if it shows already larger fluctuations in its unperturbed condition. An early example of that connection was the Sutherland--Einstein relation which gives the mobility proportional to the diffusion constant for particles suspended in an equilibrium fluid \cite{Einstein,Sutherland}.  The equilibrium fluctuation---dissipation theorem generalizes that idea and the (Green--)Kubo formul{\ae} relate the linear response coefficients of an equilibrium system to correlation functions in the unperturbed system \cite{Green1954,Kubo1957}. More precisely, they tell us that to linear order around equilibrium the response is given by a correlation of the observable with the entropy flux generated by the perturbation.  
Let us start by a newer view on that result.\\

When dynamics enters into the statistical description of physical systems it appears useful to consider dynamical ensembles \cite{leb}. On the formal side those are very reminiscent of what we more generally call path-space integration. One considers a system on a coarse-grained level with states $x$ and we denote the possible trajectories over time-interval $[0,t]$ by $\omega = (x_\tau, 0\leq \tau \leq t)$.  The states $x$ can be values of macroscopic variables or they could denote the position and momenta of some particles etc.
The state evolves in time and we can first consider an equilibrium process  where the path probabilities
are time-reversal symmetric: the probability of each trajectory satisfies
\[
\text{Prob}[\omega] = \text{Prob}[\theta\omega]
\]
where $\theta\omega$ denotes the time reversed trajectory obtained by inverting the the time-sequence of states and, if the state involves velocities, also flipping the sign of all velocities. The above relation gets often described as microscopic reversibility or detailed balance.

% 
% For time-reversal we need to invert the time-sequence of states and, if the state involves velocities, also flip the sign of all velocities.  For short  we simply denote here the time-reversed trajectory by $\theta\omega$.\\  
% Imagine now that our system has path-probabilities which are time-reversal symmetric: the probability of each trajectory satisfies 
% \[
% \text{Prob}[\omega] = \text{Prob}[\theta\omega]
% \]
% That is the case for equilibrium processes and gets often described as microscopic reversibility or detailed balance. 

We now start the system in equilibrium at time zero but then by a perturbation the system is pushed out and its statistics follows another dynamical ensemble with probabilities 
\begin{equation}\label{ac}
\text{Prob}^h[\omega] = e^{-A(\omega)}\, \text{Prob}[\omega]
\end{equation}
where the superscript $h$ stands for the amplitude of the (unspecified) perturbation and $A=A^h$ is called the action of order $h$.  We want to know the value of some path-observable $O(\omega)$ during $[0,t]$.  We can compute its expectation value as follows:
\begin{eqnarray}
\langle O(\omega)\rangle^h &=& \sum_{\omega} O(\omega)\,e^{-A(\omega)}\, \text{Prob}[\omega]\quad\text{and similarly}\nonumber\\
\langle O(\theta\omega)\rangle^h &=& \sum_{\omega} O(\theta\omega) e^{-A(\omega)}\, \text{Prob}[\omega]\nonumber\\
&=&\sum_{\omega} O(\omega) e^{-A(\theta\omega)}\,\text{Prob}[\omega]\nonumber
\end{eqnarray}
(The sums over the trajectories $\omega$ are a bit formal but that does not matter here.)
By subtracting the last line from the first line we thus obtain
\begin{equation}\label{oh}
\langle O(\omega)\rangle^h - \langle O(\theta\omega)\rangle^h = \sum_{\omega} O(\omega)\,\left[e^{-A(\omega)}-e^{-A(\theta\omega)}\right]\, \text{Prob}[\omega]
\end{equation}
Such formul{\ae} are useful for response theory.  For example, we can take for observable $O$ a particle current (where $O\theta =-O$) or even simpler we can choose $O(\omega) = \delta(x_t-x)$, the indicator to be in state $x$ at the final time $t$.  In the last case we find
\begin{equation}\label{ohh}
\text{Prob}^h[x_t=x] - \text{Prob}^h[x_0=x] = \sum_{\omega:x_t=x}\left[e^{-A(\omega)}-e^{-A(\theta\omega)}\right]\, \text{Prob}[\omega]
\end{equation}
Using that in the reference equilibrium, $\text{Prob}[x_t=x]=\text{Prob}[x_0=x]$ and since at time zero $\text{Prob}^h[x_0=x]=\text{Prob}[x_0=x]$, the perturbation is not yet visible, formula \eqref{ohh} gives the change by the perturbation as
\begin{equation}\label{coh}
\text{Prob}^h[x_t=x] - \text{Prob}[x_t=x] = \sum_{\omega:x_t=x}\left[e^{-A(\omega)}-e^{-A(\theta\omega)}\right]\, \text{Prob}[\omega]
\end{equation}
Response theory was first systematically developed some 60 years ago for dynamical ensembles that have a time-reversal symmetry and when the perturbation can be taken very small.  That is the context of linear response theory around equilibrium. While that is not the usual textbook presentation, the above can be applied immediately. We should expand the exponential to first order in $h$ (linear order perturbation theory).  Then \eqref{coh} simplifies to
\begin{equation}\label{scoh}
\text{Prob}^h[x_t=x] - \text{Prob}[x_t=x] = \sum_{\omega:x_t=x}\left[A(\theta\omega)-A(\omega)\right]\, \text{Prob}[\omega]
\end{equation}
and we see that the linear response formula around equilibrium is completely expressed by the time-reversal antisymmetric part 
$A(\theta\omega)-A(\omega)$ of the action that describes the perturbation. That is interesting because in many situations that antisymmetric part has a clear physical meaning.  That measure of time-reversal breaking is indeed related to dissipation and entropy fluxes, at least if the environment of the system remains in local thermal equilibrium.  That makes the relation between linear response around equilibrium and entropy fluxes, as summarized in the fluctuation--dissipation theorem one of the manifestations of what we can call the miracle of equilibrium: there is basically a unique entropy, having many faces in terms of different thermodynamic potentials, that governs heat exchange with the environment via Clausius theorem, appears as a time-increasing function in Boltzmann's H-theorem, characterizes macroscopic fluctuations in the system as pioneered by Planck and Einstein, specifies the statistical forces as clarified  by Onsager and, as explained above, also decides the linear response as developed by e.g. Kubo.\\

%Knowing energy and entropy (along with other macroscopic) completes teh characterization of the system in the thermodynamic sense - state functions 

We can hope for more life (but maybe less luck) if the reference system is out of equilibrium or when considering nonlinear response around equilibrium. Energy--entropy considerations only are most likely not enough to characterize nonequilibrium systems.  In fact it would be very surprising if the details of the dynamics would not matter much more for stationary nonequilibrium conditions.  Kinetic factors encoded in friction or reactivities will show up in the stationary state weights and no guarantees exist about a purely thermodynamic meaning of fluctuation or variational functionals.  This immediately implies that nonequilibrium theory cannot just be nonequilibrium thermodynamics and in general one has to go beyond the concept of entropy and work to construct nonequilibrium statistical mechanics.
%This  is already true for relaxation to 

%The problem of nonequilibrium statistical mechanics is to look for unifying concepts and physical principles 

Indeed, \eqref{scoh} will be false and the physics of the time-symmetric part in the action $A$ comes to the foreground. That corresponds to a different physics; when a perturbation is applied around a nonequilibrium condition, kinetic factors, apart from entropy, start to play a key role \cite{fdr,update}. Similarly for higher order response around equilibrium, these kinetic factors appear explicitly \cite{second}.  By `kinetic factors' we generically refer to the non-thermodynamic aspects like exact details of couplings of a system with a reservoir or time--symmetric Arrhenius prefactors in transition rates --- quantities which are irrelevant in the study of static equilibrium linear response. {\it A priori}, this seems a formidable scenario --- there could be many possible ways to perturb a system which are different kinetically but equivalent thermodynamically. Is there a generic method to describe response theory beyond the equilibrium fluctuation-dissipation relation? Luckily, it turns out that the answer is yes. The notion of `frenesy' or dynamical activity, which is a time-symmetric quantity, allows us to provide a unified framework to study nonequilibrium response \cite{fdr}.\\

\begin{figure}[ht]
 \centering
 \includegraphics[width=7 cm]{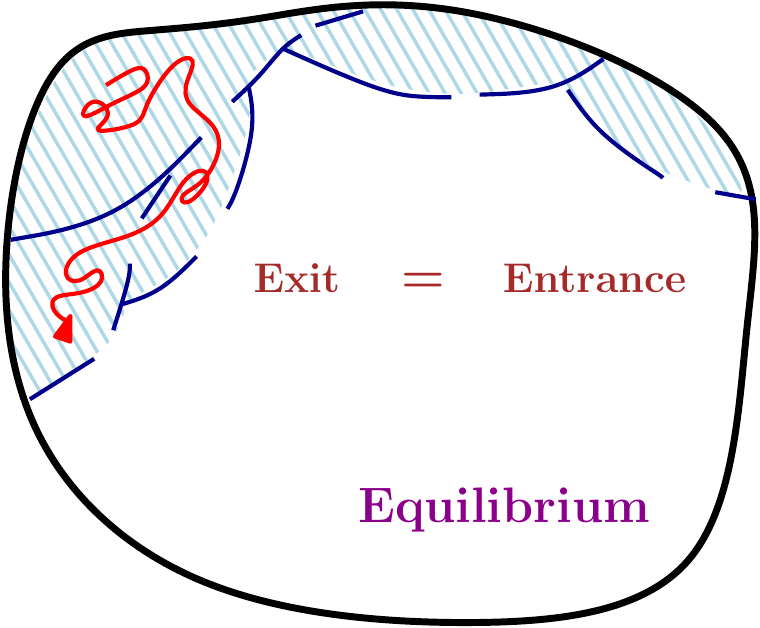}
 % ph_sp.pdf: 595x842 pixel, 72dpi, 20.99x29.70 cm, bb=0 0 595 842
 \caption{Schematic representation of the phase space of a macroscopic system. In equilibrium, which corresponds to the largest region, only volume (entropy) is relevant while surface effects (reactivity and escape rate) are substantial when the system is in a smaller region corresponding to some nonequilibrium stationary state. Reactivities are symmetric, suggested here by having exits equal to entrances between any two regions.
 }
 \label{fig:phase}
\end{figure}

Frenetic contributions have already been described in studies of nonequilibrium system in various contexts. It is the key quantity leading to negative differential mobility for particle transport in disordered systems or constrained geometries \cite{Alessandro, fren}. The dynamical activity is also an important player in studies of kinetically constrained models \cite{kcm} or glass transitions \cite{Speck}. Experimental studies of micro-swimmers also pick up the signature of this time--symmetric quantity \cite{Bechinger}. On a more theoretical level we see the frenetic contribution as a complement to entropic considerations. The latter ultimately refer to estimates on phase space volumes for a physical coarse-graining of the energy surface corresponding to the appropriate closed and isolated 
universe containing our system.  Volumes in phase space are especially important if the differences are large, with the equilibrium condition taking in fact almost all volume in phase space (see Fig. \ref{fig:phase}).  When well away from that largest volume, it does not seem unreasonable to think that also ``surfaces'' of phase space regions get important in the sense that the exit and entrance rates of the regions start much more there to influence
%determining 
response and fluctuations.  
That time-symmetric traffic between different conditions (phase space regions) is what we have in mind with dynamical activity and the frenetic contribution manifests itself via correlation functions with such a traffic or dynamical activity.\\

The aim of this article is not to provide a comprehensive review of these studies, rather to illustrate how frenetic aspects lead to some interesting phenomena with the help of some simple examples. Relevant references are provided for more detailed studies, wherever necessary. In the next section we give a brief 
account of the theoretical background of response away from equilibrium; the relevant response formul{\ae}, derived from a path-ensemble formalism are quoted. The precise meaning and definition of dynamical activity 
is also illustrated in the context of Markov jump processes.   In Sec. \ref{sec:ex} we move on to a few chosen examples starting with the toy model of a biased random walker. How negative differential mobilities can arise in presence of `trapping' in the system is discussed next. Section \ref{sec:sec} focuses  on higher order response around equilibrium which picks up kinetic details in the coupling with the reservoir. The last example in Sec. \ref{sec:popn} explores the effect of population inversion on energy transport in a simple setting of a multilane stochastic particle system. The population inversion is again caused by a suitable kinetic aspect. We conclude with some general remarks about nonequilibrium response theory.

%\textbf{more plan of paper} 

%In this article we discuss several examples where one can illustrate frenetic aspects of response using a few chosen simple models of stochastic Markov processes. 

%%%%%%%%%%%%%%%%%%%%%%%%%%%%%%%%%%%%%%%%%%%%%%%%%%%%%%%%%
\section{Response formulae}\label{sec:form}

In this section we give a brief review of the relevant response formul{\ae}. That is meant to establish the notation and to explain the concepts rather than to prove them. We refrain from giving the derivations but the references will point to more reviews.

\begin{figure}[ht]
 \centering
 \includegraphics[width=8 cm]{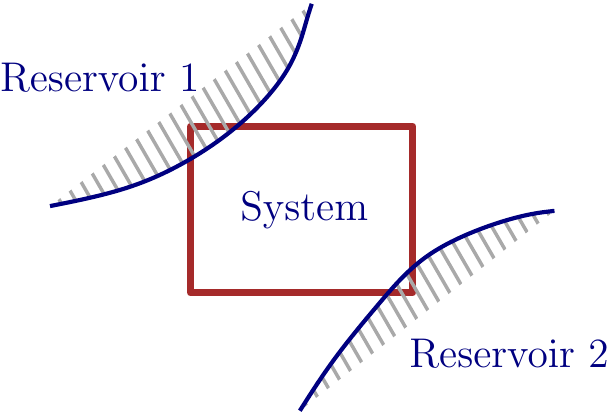}
 % set_up.pdf: 595x842 pixel, 72dpi, 20.99x29.70 cm, bb=0 0 595 842
 \caption{An open system in contact with spatially separated equilibrium reservoirs satisfies local detailed balance.}
 \label{fig:setup}
\end{figure}

\noindent  \eqref{ac} makes sense also outside equilibrium, i.e., for a reference dynamical ensemble that is not time-reversal symmetric. In a way \eqref{ac} defines what are the possible perturbations that we can consider via the path-space formalism. For the physics we move now to the context of open systems in contact with multiple spatially well-separated equilibrium reservoirs (see Fig. \ref{fig:setup} ). For systematics it is then useful to decompose the action into its time-antisymmetric and time-symmetric components,
\[
S_h : =  A\theta - A,\quad D_h := \frac 1{2}(A + A\theta), \quad A = D_h - \frac 12 S_h
\]
They come with a physical interpretation.  That is most clear and familiar for 
\[
S_h(\omega) = \log \frac{\text{Prob}^h[\omega]}{\text{Prob}^h[\theta\omega]}  - \log \frac{\text{Prob}[\omega]}{\text{Prob}[\theta\omega]} 
\]
giving a measure of the excess in time-reversal breaking by applying the perturbation.  When the system is in weak contact with spatially separated equilibrium reservoirs, that time-antisymmetric part can be identified with the entropy flux (per $k_B$) into that environment along the path $\omega$.  It is in other words a thermodynamic quantity related to heat and dissipation of work, and the reason is summarized as the condition of local detailed balance; see \cite{leb,kls,time,har,der,hal}. Also the dynamical activity $D_h(\omega)$ is identified for each trajectory $\omega$; it is time-symmetric and its role and meaning is part of the questions in today's attempts towards constructing nonequilibrium statistical mechanics.  Also this paper will contribute to illustrating its nature. For being more specific  we  illustrate below, using the example of a Markov jump process, how these quantities can be computed for a given dynamical ensemble \cite{fdr,update,fren}. 

\subsection{Linear response around nonequilibrium}
Formula \eqref{ac} implies that $\langle O(\omega)\rangle^h = \langle e^{-A(\omega)}\,O(\omega)\rangle^0$.  That immediately gives the linear differential response to the perturbation,
\bea
\frac \id{\id h} \la O(\omega)\ra^h = \frac 12 \left \la O(\omega); S_h'(\omega) \right\ra^h - \left \la O(\omega); D_h'(\omega)\right\ra^h ~~~\label{eq:dQSD_cov}
\eea
$\la f;g \ra$ denotes the covariance between the observables $f,g$ and $S_h',D_h'$ are first derivatives with respect to the drive $h.$  The averages $\langle\cdot\rangle^h$ are over trajectories including possibly the initial conditions and depending on the driving field $h$.  We will often drop the explicit dependence on $h$ in the notation.   Thus, the first term in \eqref{eq:dQSD_cov} signifies the covariance or the connected correlation of the observable $O$ with the linear excess of entropy generated due to the perturbation and the second term arises from the correlation with the change in dynamical activity.  Note also from \eqref{eq:dQSD_cov}
that whenever $O$ only depends on the initial state $x_0$, then the left-hand side is zero and so must the right-hand side
\begin{equation}\label{eqq}
\frac 12 \left \la O(x_0); S_h'(\omega) \right\ra^h = \left \la O(x_0); D_h'(\omega)\right\ra^h
\end{equation}
which relates the entropy flux with the dynamical activity.  Also observe that we can take as a special case $O(\omega) = S_0'(\omega)$ in \eqref{eq:dQSD_cov}, to find
\bea
\frac \id{\id h} \la  S_0'(\omega)\ra^h = \frac 12 \left \la S_0'(\omega); S_0'(\omega) \right\ra^h - \left \la S_0'(\omega); D_h'(\omega)\right\ra^h ~~~\label{eqS}
\eea
where the first term is always positive.  In fact, the observable $S_0'(\omega)$ is time-antisymmetric and refers to an excess of entropy flux carried by the different currents in the system.
The formula \eqref{eqS} is therefore an extended Green-Kubo formula (in the Helfand form) for the differential transport coefficients. We remark also that from \eqref{eq:dQSD_cov}-\eqref{eqS} we can obtain
the Harada-Sasa equality connecting energy dissipation with the violation of the usual fluctuation--response relation, \cite{har,maral}.  Finally we feel that it makes sense to make a connection here with the notion of effective temperature.  Effective temperature from the response formula \eqref{eq:dQSD_cov} can be defined by dividing that formula by $\left \la O(\omega); S_h'(\omega) \right\ra^h$.  In the case of equilibrium that would give the (real thermodynamic) temperature; for nonequilibrium, there is a correction involving the frenetic contribution.\\

Let us look again at the special case of perturbing around equilibrium  $h=0.$
%(also including an initial averaging over the equilibrium distribution) and additionally 
In that case, if the observable $O$ is antisymmetric under time-reversal, e.g. being a time-integrated energy or particle current $O(\omega)=J(\omega)$, then
$\left \la J(\omega) ; D_0'(\omega) \right\ra^0 =0$ because the dynamical activity $D_h(\omega)$ in \eqref{eq:dQSD_cov} is itself time-symmetric and the equilibrium process is time-reversal invariant.  Thence,
\bea
\left.\frac \id{\id h} \la J(\omega)\ra^h\right |_{h=0} = \frac 12 \left \la J(\omega) ;  S_0'(\omega) \right\ra^0 \label{eq:equilbis}
\eea
That equilibrium result is  the well--known Green--Kubo relation for equilibrium conductivities. We will see it more explicitly in later examples. Another special case is when the observable is a state function $O(\omega)=O(x_t).$  Then, in equilibrium, \eqref{eq:dQSD_cov} reduces to the Kubo formula,
\bea
\left.\frac \id{\id h} \la O(x_t)\ra^h\right |_{h=0} =  \left \la O(x_t) ;  S_0'(\omega) \right\ra^0 = \left \la O(x_t)   S_0'(\omega) \right\ra^0\label{eq:equilbkubo}
\eea
The first equality follows from
\[
\langle O(x_t)D'_0(\omega)\rangle^0 = \langle O(x_0)D'_0(\omega)\rangle^0 = \frac 1{2}\langle O(x_0)S'_0(\omega)\rangle^0 = -\frac 1{2}\langle O(x_t)S'_0(\omega)\rangle^0 
\]
where we have used \eqref{eqq} for the middle equality. All the other equalities follow by time-reversal invariance of the equilibrium process $\langle \cdot\rangle^0$.\\

For the response in \eqref{eq:equilbis}--\eqref{eq:equilbkubo}
around equilibrium it suffices to know how the perturbation affects the entropy flux (or, excess dissipation in the environment). Away from equilibrium, however, the frenetic contribution $\left \la O(\omega) ;  D_h'(\omega) \right\ra^h$ in \eqref{eq:dQSD_cov} involving the dynamical activity $D(\omega)$ plays a major role. As we will see later, this $D(\omega)$ explicitly involves `kinetic factors' like escape rates, symmetric prefactors or diffusion constants, which implies that  the nonequilibrium response formula carries more detailed information about the system than the equilibrium one.

 %In particular, a large frenetic contribution  can also result in a negative differential response $\frac d{dh} \la O(\omega)\ra^h \leq 0$ in some regime of the parameter $h,$ even in cases where that is strictly forbidden and not possible in equilibrium \cite{fren}.
 
 \subsection{Higher order response}
 
Using dynamical ensembles allows one also to easily investigate higher order responses. In particular, to second order  the general nonequilibrium response formula takes the form
\bea
\frac {\id^2}{\id h^2} \la O(\omega)\ra^h &=& - \la D_h''(\omega) O(\omega) \ra^h + \la (D_h'(\omega))^2 O(\omega) \ra^h \cr
&& +\frac 14  \la (S_h'(\omega))^2 O(\omega)\ra^h - \la S_h'(\omega) D_h'(\omega) O(\omega)\ra^h \label{eq:sec}
\eea

 Here we have assumed that $S_h''=0$, i.e., the entropy flux $S_h$ is linear in the field $h$ which is true for most physically relevant examples of perturbations.  Unsurprisingly, the dynamical activity appears explicitly in this nonequilibrium second order response formula. What is interesting is that the frenetic effects become important even when one looks at response around equilibrium from second order onwards. In particular for a time-antisymmetric observable $J(\omega),$ and the perturbation around  equilibrium $h=0$  one obtains the second order correction to the Green-Kubo formula \cite{second},
 \bea
 \left. \frac {\id^2}{\id h^2} \la J(\omega)\ra^h \right |_{h=0} = - \la S_0'(\omega) D_0'(\omega) J(\omega) \ra^0 \label{eq:secJ_eq}
 \eea
 The extension of the Kubo formula for a state observable $O(x_t)$ is the second order response around equilibrium,
 \bea
 \left. \frac {\id^2}{\id h^2} \la O(x_t)\ra^h \right |_{h=0} = -2 \la S_0'(\omega) D_0'(\omega) O(x_t)\ra^0 \label{eq:sec_eq}
 \eea
 Indeed the above expressions depend explicitly on the  excess in  the dynamical activity $D_0'$ generated by the perturbation while the first order, as given by \eqref{eq:equilbkubo}, does not. The direct implication is that perturbations which are thermodynamically equivalent --- generating the same entropy fluxes but differing in kinetic details, would have different second order responses.\\

 %Later we will illustrate with  examples.  

 Before we discuss the frenetic aspects of response in more details with the help of several physically motivated examples 
 %and see how they can be used to 
 let us make more explicit  the quantities $S_h(\omega)$ and  $D_h(\omega)$  for systems modelled by Markov jump processes. 
 We consider a Markov jump process specified by transition rates $k(x,y)$ for jumps $x\rightarrow y$ between states $x,y$.  
 These are parametrized by the time-antisymmetric and symmetric $s(x,y)$ and $\psi(x,y)$ respectively
\bea
k(x,y) &=& \psi(x,y)\,e^{s(x,y)/2},\cr
\psi(x,y)&=&\psi(y,x)\geq 0, \; s(x,y)= -s(y,x) \label{eq:psi_s}
\eea
all possibly depending on the field or potential with amplitude  $h$.   A trajectory $\omega:=(x_\tau, 0\le \tau \le t)$ over time interval $[0,t]$  is characterized by discrete jumps at times $\tau_i$ and by exponentially distributed waiting times $\tau_{i+1}-\tau_i.$ The entropy flux and dynamical activity associated with this trajectory (up to some additive constants which depends on the reference process) are
\bea
S_h(\omega) &=&  \sum_i s(x_{\tau_i},x_{\tau_{i+1}}) \cr
D_h(\omega) &=& \int_0^t \id \tau \, \xi(x_\tau) - \sum_{i} \log \psi(x_{\tau_i},x_{\tau_{i+1}})  \label{eq:ShDh}
\eea
where $\xi(x) = \sum_y k(x,y)$ is the escape rate at state $x$.  
The dynamical activity is time-symmetric; reversing the time over the trajectory in $[0,t]$ does not change it.  It consists of two contributions, the reactivities $\psi$ and the escape rates $\xi$. These are the  {\it kinetic factors} that become particularly significant away from equilibrium. One of the interesting questions is to discover how $\psi$ is influenced by the nonequilibrium condition and how it changes with $h$. The response will teach us about some of that dependence. On the other hand, $S_h$ is time-antisymmetric and corresponds to the {\it thermodynamic} entropy flux over $[0,t]$ when the condition of local detailed balance is satisfied.  Then $s(x,y)$ denotes the entropy flux to the environment (per $k_B$) in the transition $x\rightarrow y$. The excess entropy and dynamical activity produced by the perturbation is calculated from the above Eq. \eqref{eq:ShDh} and must  be used in the relevant response formulae \eqref{eq:dQSD_cov}, \eqref{eq:sec} or \eqref{eq:sec_eq}.  Note that under (global) detailed balance $s(x,y) = \beta [U(x)-U(y)]$ for some potential $U$, and then the stationary equilibrium distribution $\sim \exp[-\beta U]$ is completely independent of the prefactors $\psi(x,y)$.

\section{Examples}\label{sec:ex}
In this section we discuss several examples where kinetic factors play an important role in the response behaviour.

\subsection{Toy model: 1D Random walk }\label{sec:rw}

\begin{figure}[th]
 \centering \includegraphics[width=8 cm]{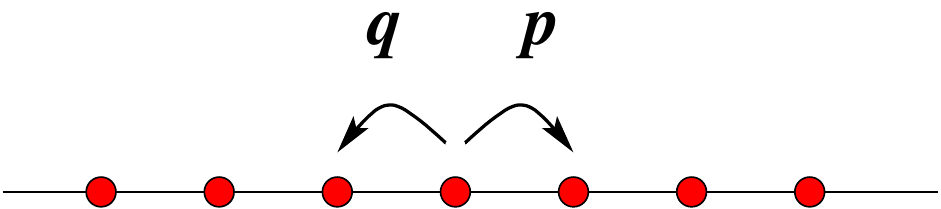}
 % random.jpg: 367x69 pixel, 72dpi, 12.95x2.43 cm, bb=0 0 367 69
 \caption{The usual parametrization for a continuous time biased  random walk.}
 \label{fig:rwalk}
\end{figure}

Our first example is a toy model of a biased one-dimensional random walk on a lattice which, despite being extremely simple, serves as a paradigmatic example to illustrate the notions of kinetic factors and frenetic contributions appearing in nonequilibrium response.\\
Let us consider a one-dimensional nearest neighbour continuous time random walk specified by  rates $p$ and $q$ of jumping to the right, respectively left neighbour (see Fig. \ref{fig:rwalk}). In the parametrization \eqref{eq:psi_s},
\[
\psi(x,x \pm 1) = \sqrt{pq},\quad s(x,x\pm 1) = \pm \log \frac{p}{q},\quad x\in \bbZ
\]

Equilibrium dynamics corresponds to $p=q$. Away from equilibrium, the local detailed balance condition characterizes the ratio of rates,
\bea
\log \frac{p}{q} = \beta E \n
\eea
for an environment at temperature $\beta^{-1}$ and with external field $E$ to the right. The symmetric prefactor or the reactivity $\psi$ however is not specified yet by that condition. In general, one can imagine that $\psi = \psi_\beta(E)$ is also affected by the field $E;$ the exact functional form would depend on the specific way the system is coupled to the environment and the driving agency.  The walker is just an effective or reduced description of a possibly much more complicated process.\\

\begin{figure}[ht]
 \centering
 \includegraphics[width=14 cm]{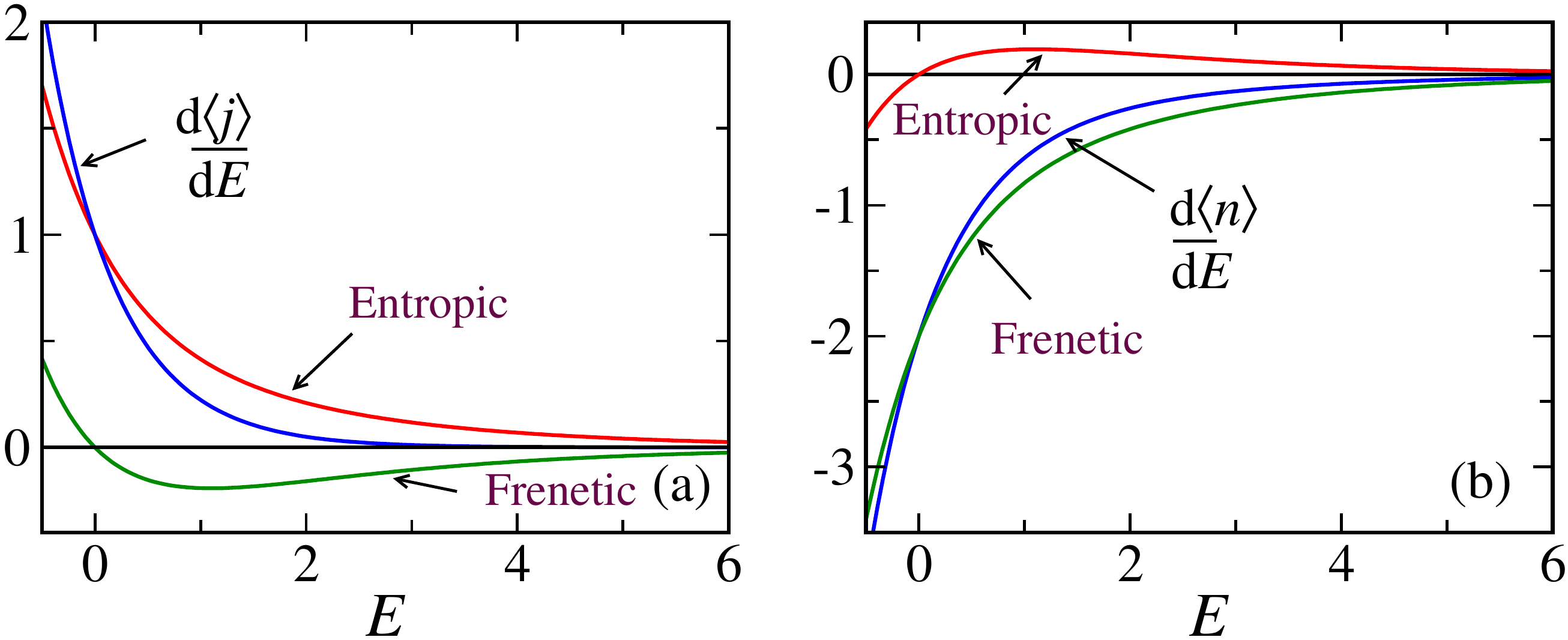}
 % J_N_rw.pdf: 780x299 pixel, 72dpi, 27.52x10.55 cm, bb=0 0 780 299
 \caption{Linear response for 1-dimensional biased random walk. (a) The differential conductivity $\frac{\id \la j \ra}{\id E}$ along with its entropic and frenetic components are plotted as a function of the bias $E.$ (b) The same for the time-symmetric traffic $\la n \ra.$
 Here $\psi_\beta(E) = \exp(-\beta E).$}
 \label{fig:rw_JN}
\end{figure}

We look now at the differential response when the drive $E \to E+\id E.$ It is needless to say that for this simple random walk model all the physical properties can be calculated exactly, and one does not require linear response theory to compute the effect of the perturbation. Again, this biased random walker acts as an effective model for studying negative differential response in  particle transport. \\
The entropy flux and dynamical activity for a given trajectory $\omega$ are given by \eqref{eq:ShDh},
\bea
S(\omega) &=& (N_+-N_-) \log \frac pq = \beta E J \cr
D(\omega) &=& (p+q)t -\frac 12 (N_++N_-) \log pq  \cr
         &=& 2\psi_\beta(E) \cosh \frac{\beta E}2 t - N \log \psi_\beta(E)\n
\eea
where $N_+$ and $N_-$ denote the number of jumps in $[0,t]$ to the right and the left respectively.  $J = N_+-N_-$ is the time-integrated current along the field and  $N=N_++N_-$ is the  `traffic' $i.e.$ the total number of jumps during the interval $[0,t].$ The primary observable of interest is the current $J;$ the differential conductivity  is then predicted using \eqref{eq:dQSD_cov} and writing $h=E,$
\bea
{\id \over \id E}\la J\ra &=& \frac 12   \left \la J; S'(\omega)\right \ra - \left \la J; D'(\omega) \right \ra \cr
&=& \frac \beta 2 \la J;J \ra + \frac{\psi_\beta'(E)}{\psi_\beta(E)} \la N;J\ra~  \label{eq:dJdE_rw} \\
&=& \left[\beta \psi_\beta(E)  \cosh \frac{\beta E}2 + 2 \psi_\beta'(E) \sinh \frac{\beta E}2 \right]t \label{eq:dJdE_psi} 
\eea

The first term in \eqref{eq:dJdE_rw}, the variance of the current $J$, is the positive entropic contribution.  The correlation $\la N;J\ra$ arises from the reactivity and note that it vanishes in equilibrium $(E=0)$.  In other words the change in the  prefactor $\psi_\beta(E)$ as function of $E$ remains unnoticed in linear response around equilibrium. The interesting scenario is when the second term contributes sufficiently negatively yielding a negative differential response.  From  \eqref{eq:dJdE_psi} we see what is needed --- $\psi_\beta'(E)$ should be `sufficiently' negative. 
Fig. \ref{fig:rw_JN}(a) shows the differential conductivity for such a choice of $\psi_\beta(E) = \exp(-\beta E).$ For the sake of convenience we have used the current per unit time $j=J/t$ as the observable; the entropic and frenetic contributions are also shown separately. %We will see more physical examples  of 

Similarly, one can also look for the response in the traffic $N = N_+ +N_-,$
\bea
{\id \over \id E}\la N\ra &=& \frac \beta 2 \left \la N; J\right \ra + \frac{\psi_\beta'(E)}{\psi_\beta(E)}  \left \la N; N \right \ra \cr
&=& \left[\beta \psi_\beta(E)  \sinh \frac{\beta E}2 + 2 \psi_\beta'(E) \cosh \frac{\beta E}2 \right]t 
\eea
Note that there the variance of $N$ appears in the frenetic term. That was seen already in the context of linear response for the boundary driven zero range process \cite{Alberto}.  In fact, since $N$ is time-symmetric the role of entropic and frenetic contribution gets reversed; even for $E=0$ we get only the frenetic part. That is reminiscent of the situation for the momentum current in shear experiments.  The viscosity is then in fact frenetic and not entropic. That similarity has been discussed in e.g. \cite{Sellitto}.\\
 Fig. \ref{fig:rw_JN}(b) shows the response and the entropic and frenetic components separately for $\la n\ra= \la N\ra /t,$ the average number of undirected jumps per unit time. We see that the differential response for $\la n\ra$ is negative for this specific choice of $\psi(E).$

The second order response for the current $J$ around equilibrium also immediately follows from Eq. \eqref{eq:secJ_eq},
\bea
\left. \frac {\id^2}{\id E^2} \la J(\omega)\ra \right |_{E=0} &=& -\la S_0'(\omega) D_0'(\omega) J(\omega)\ra^0 \cr
&=& - \beta \left[2 \psi_\beta'(0) t  \la J^2 \ra^0 - \frac{\psi_\beta'(0) }{\psi_\beta(0)} \la J^2 N \ra^0 \right]\cr
&=& 2 \beta \psi_\beta'(0) t \n
\eea
Clearly that second order response now depends explicitly on  how the reactivity $\psi$ changes with the driving $E$. In particular, as already mentioned in the previous section, perturbations which are equivalent thermodynamically $i.e.$, produce the same entropy fluxes but differ in their effect upon the kinetic prefactor would generate separate second order responses. 
%To put it the other way, second order response provides more detailed information about the system.

\subsection{Trapping effect: negative differential mobility}\label{sec:trap}

Mobility of a particle quantifies how its velocity changes when subjected to an external mechanical force. 
Requirements for thermodynamic stability ensure that the mobility is strictly positive around equilibrium 
$i.e$ the velocity or current increases when the applied force is larger. This is a consequence of the fact that the equilibrium mobility, as given by the Sutherland-Einstein relation,  is nothing but the variance of the current in the unperturbed state. This statement is no longer true when the system is driven far away from equilibrium, where the frenetic contribution becomes relevant, \cite{prs}. There one speaks about the differential mobility  which characterizes how the current changes with the drive. One of the consequences of the modified Kubo formula is that negative differential mobilities are allowed to exist, $i.e.$, the current need not be a monotonically increasing function of the drive.  \\

\begin{figure}[ht]
 \centering
 \includegraphics[width=12 cm]{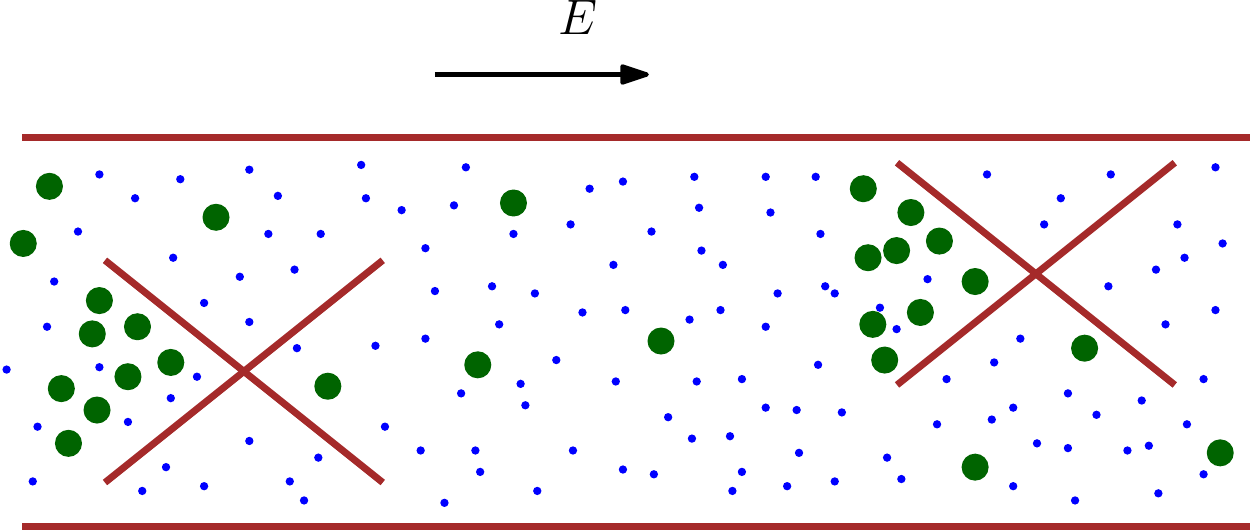}
 % trap1.pdf: 595x842 pixel, 72dpi, 20.99x29.70 cm, bb=0 0 595 842
 \caption{Driven particle motion in constrained geometry giving rise to trapping.  Colloid particles (dark green discs) tend to reside inside the cages created by the obstacles when pushed hard.
 The blue dots represent the equilibrium medium. } 
 \label{fig:trap}
\end{figure}

Several studies of negative differential mobility in nonequilibrium  particle systems can be found in the literature. Examples include particles diffusing in restrictive or confined geometries \cite{fren,Zia, thesisSOghra}, driven lattice Lorentz gas \cite{Barma,Franosch}, driven particle in an environment of mobile obstacles \cite{Alessandro, mobility} and exclusion processes with constrained dynamics \cite{Sellitto}. \\
To understand the origin of such negative differential mobility from a general viewpoint let us look at Eq. \eqref{eq:dQSD_cov} where $O=J,$ the time-integrated current, in what we imagine as a driven particle system. The excess entropy is essentially proportional to the current itself implying the positivity of the entropic term. A large negative contribution from the frenetic term, however, could result in an overall negative differential response of the current. \\

It was shown in \cite{fren,Zia} that this indeed might happen if there is a possibility of trapping in the system. Intuitively that is quite easy to understand --- if the particles get trapped then the dynamical activity is affected and if it so happens that the trapping or caging increases when the perturbing field becomes stronger then negative differential mobility can occur.  A simple example is the driven motion of colloids in a narrow channel in presence of directed `cages' or obstacles (Fig. \ref{fig:trap}); the particles get more and more trapped when the driving force is increased resulting in a decrease of the current. It is important, however, to note that `trapping' need not always mean a real trap or cage in the physical space, it might as well refer to effective trapping in some region of the phase space (cf. Fig. \ref{fig:phase}).

The biased random walk discussed in the previous section can be seen as a paradigmatic example of this phenomenon --- if the driving affects the reactivities in such a way that the escape rate is a decreasing function of the drive, then the current shows a non-monotonic behaviour. This was discussed in detail in \cite{Alessandro,fren}, where it was shown that indeed the frenetic component is responsible for the negative differential mobility. Here it suffices to mention that several models of interacting particle systems can effectively be mapped to this simple picture of biased random walk.

\subsection{Coupling with reservoir}\label{sec:sec}

Our next example illustrates how the second order response picks up  kinetic details of the coupling to a particle reservoir.\\
We consider a tracer particle of mass $M$ with position $x_t$ coupled to an equilibrium bath, in the sense that it obeys the Newton equation
\begin{equation}\label{newton}
M\ddot x_t = -\nabla_x U(x_t,\eta_t) =: F(x_t,\eta_t)
\end{equation}
where the $\eta$ denotes a family of $N$  degrees of freedom making the equilibrium reservoir. $\eta$ would typically contain position $r_i$ and velocity $\dot{r}_i$ of the reservoir particles. The force $F(x_t,\eta_t)$ acting on the particle is ``noisy'' as the $\eta_t$ gets  sampled from the equilibrium process which in back-reaction depends on the history $(x_s,\, s \leq t)$ of the particle.
  
The bath degrees of freedom evolve under a dynamics  where each position $r_i, i=1,\ldots,N$ is coupled to the probe via the same $U(x,\eta)$ and possibly with other interactions while on the whole satisfying detailed balance at fixed temperature $\beta^{-1}$. For greater simplicity we take a coupling
(in one-dimensional notation)
\begin{equation}
U(x,\eta) = \frac{\lambda}{2} \sum_i (x-u_i(\eta))^2,\quad F(x,\eta) = \lambda\sum_i (u_i(\eta)-x)
\end{equation}
so that for some function $u_i(\eta)$
\begin{eqnarray}
-\frac{\partial}{\partial r_j}U(x,\eta) &=& -\lambda\sum_i u_i(\eta)\frac{\partial}{\partial r_j}u_i(\eta) + x\,\frac{\partial}{\partial r_j}V(\eta)\nonumber\\
V(\eta) &:=& \lambda\sum_i u_i(\eta)
\end{eqnarray}
That choice implies that for the $\eta-$dynamics, changing the position $x\rightarrow x+\id x$ means simply to add the potential $V$ with small amplitude $\id x$
with respect to its original dynamics.  The response below will be written in terms of that perturbing potential $V$.  Note that $0<\lambda\sim 1/N$ to mimic a mean field coupling in which the particle couples equally with all the bath degrees of freedom.\\

Integrating out the equilibrium bath from \eqref{newton} should give rise to an effective and generalized Langevin dynamics for the (tracer) particle,
%
%The evolution \eqref{newton} is written as
\begin{equation}\label{newton1}
M\,\ddot x_t =  -\lambda Nx +\langle V(\eta_t) \rangle^{(x_s, s\leq t)} + \xi_t
\end{equation}
with the noise
$\xi_t := V(\eta_t) - \langle V(\eta_t) \rangle^{(x_s, s\leq t)}$ by construction dependent on the whole history
$(x_s,\,s\leq t)$.  The expectation $\langle \cdot\rangle^{(x_s, s\leq t)}$ is in the $\eta-$process conditioned on a  fixed particle trajectory up to time $t$.  In other words, we treat the particle trajectory as a fixed protocol in the dynamics for $\eta$.   The regime where our perturbation approach makes sense is when there occurs
 a time-scale separation between the characteristic times
$\tau_x$ of the particle and $\tau_\eta$ of the sea.  To evaluate the expectation in \eqref{newton1} perturbatively we take as reference the stationary $\eta-$dynamics with fixed $x_s\equiv x_t, s\leq t$, the particle position at time $t$. In other words we have a time-dependent perturbation amplitude $h_s = x_s-x_t$.  We apply the response formalism  of \cite{second} to second order,
\begin{eqnarray}
\langle V(\eta_t) \rangle^{(x_s,s\leq t)} &=& \langle V \rangle_{x_t}\nonumber\\
&+&\beta\, \int_{-\infty}^t \frac{\id}{\id s} \langle V(\eta_s)\,V(\eta_t)\rangle_{x_t}\cdot (x_s - x_t)\,\id s \nonumber\\
&+& \frac{\beta^2}{2} \int_{-\infty}^t \id s \int_{-\infty}^t \id s'\, \langle V(\eta_s)\, L_{x_t}V(\eta_{s'})\,V(\eta_t)\rangle_{x_t} \, \dot{x}_s\,(x_{s'}-x_t)\label{fr2}
\end{eqnarray}
where $L_{x_t}$ is the backward generator of the $\eta$-equilibrium process at fixed particle position $x_t$. \\
The first line contains the time-local term  which is the statistical force in the infinite time-scale separation limit between bath and particle. The expectation $\langle \cdot\rangle_x$ is in the stationary equilibrium for the reservoir at fixed particle position $x$.  That statistical force will be given by the gradient of the equilibrium free energy constrained to $x$.    The second line arises from the standard Kubo-term in linear response.  It gives rise, via partial integration, to the usual friction which is proportional to the noise covariance (what gives the second fluctuation--dissipation relation, at least in the absence of the last line).  The last line indeed gives the correction coming from second order (nonlinear) response, containing frenetic aspects.  The point is that $L_x$ generates the time-evolution of the reservoir (at fixed particle position $x$) and thus contains the microscopic interaction between the reservoir degrees of freedom.\\
Equation \eqref{fr2} must be substituted into \eqref{newton1}
for obtaining the Langevin-type effective dynamics of the particle.  The reservoir kinetics sits already in the time-correlation functions that make the usual friction and noise term, but the last line in \eqref{fr2} adds significantly to that.  The details of the dynamics will enter explicitly even in the Markov approximation.  To the best of our knowledge that additional frenetic aspect has not been studied before, but we hope it will be experimentally visible. A similar study can be done when starting with nonequlibrium reservoirs. We then need to apply formula \eqref{eq:dQSD_cov} for obtaining the friction; see \cite{chr,stefano}.

\noindent We conclude this discussion with the remark that  if the bath is linear, say consisting of harmonic oscillators like in Section 1.6 of Zwanzig's book, \cite{zwanzig}, we can do an exact calculation with independent bath particles.  Then the linear (Kubo) order for response suffices and there are in fact no corrections of second order.  Therefore, saying it differently, forgetting about second order (kinetic) features in deriving the Langevin equation is like forgetting nonharmonic effects in the equilibrium bath.\\

\subsection{Population inversion}\label{sec:popn}

Our last example centres around another typical nonequilibrium phenomenon, namely population inversion. Population inversion is about being able to select a certain (output) stationary statistical distribution over the various available states of the system.  In equilibrium, temperature and energy decide and we are stuck with the Maxwell-Boltzmann statistics.  Driving conditions add flexibility.  It is not however the driving itself or only the driving that decides the (new) stationary distribution ---  the reactivities or time-symmetric factors in the transition rates are (also) essential.  A very simple example is to take a ring with $L$ sites and a biased walker with rates $k(x,x +1) = a_x e^{E/2}, k(x+1,x) = a_x e^{-E/2}$.  For $a_x\equiv 1$ the stationary distribution is uniform for no matter what $E$.  But for large $E$ we can basically concentrate the stationary distribution on whatever site $x$ we want by a suitable choice of $a_x$.  We believe that such a mechanism of kinetic selection is widespread in bio-chemical systems for better efficiency in getting the right output product. Since that mechanism is time-symmetric and works (only) out-of-equilibrium we call it again frenetic.  Similarly, for the origin of laser working one needd to produce a population inversion.
It is well--known that at low temperatures, a suitable choice of symmetric prefactors can select  any desired (even high) energy level of a multilevel system \cite{Winny}, so as to establish an inversion of the population with respect to the usual Boltzmann statistics. Next we investigate how this population inversion affects the energy transport in a spatially extended system.  Such energy transport can now show unusual features which ultimately go back to frenetic aspects playing a role in the energy reservoirs.   In fact we will drive one energy bath out-of-equilibrium and enforce in it a population inversion.\\

\begin{figure}[th]
 \centering
 \includegraphics[width=10 cm]{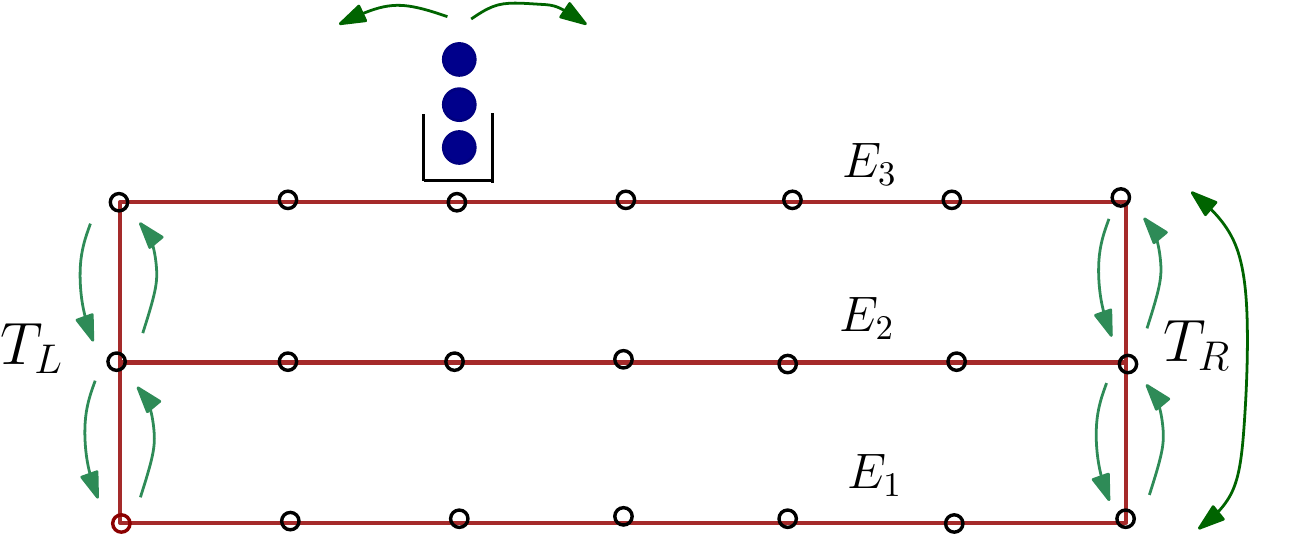} 
 % mult_zrp.pdf: 595x842 pixel, 72dpi, 20.99x29.70 cm, bb=0 0 595 842
 \caption{Schematic representation of the multilane particle transport model coupled to two thermal reservoirs (with $K=3$).}
 \label{fig:zrp}
\end{figure}

The model is defined on a two--dimensional lattice; the horizontal direction is interpreted as space and the vertical one as energy.  The different energy levels are spatially invariant and  we refer to them as lanes. Following this convention a $L \times K$ lattice corresponds to a spatially extended one--dimensional system with $L$ sites, each site having $K$ possible energy levels having energies $E_1 < E_2 \dots < E_K$.
Each site on this lattice can accommodate any number of particles; the configuration of the system is fully specified by the set $\{ n_{i\ell}; i \in [L], \ell \in [K]) \}$ where $n_{i\ell}$ is the number of particles at the $\ell^{th}$ level of the $i^{th}$ site. We suppose that energy is conserved in the bulk, $i.e.$, particles cannot change lanes there; it can only hop to its neighboring sites on the same lane with some rate which we take to be unity here. A particle can jump across the energy lane at the left and right boundaries where the system is connected to two  thermal reservoirs with temperatures $T_L$ and $T_R$ respectively (see Fig. \ref{fig:zrp}); the system can gain or lose energy when a particle changes the lane. We assume free particle motion at the boundary sites so that the jump rate is proportional to the number of particles $n$ at the source site. (The detailed dynamics is specified later.)\\

\begin{figure}[th]
 \centering
\includegraphics[width=14 cm]{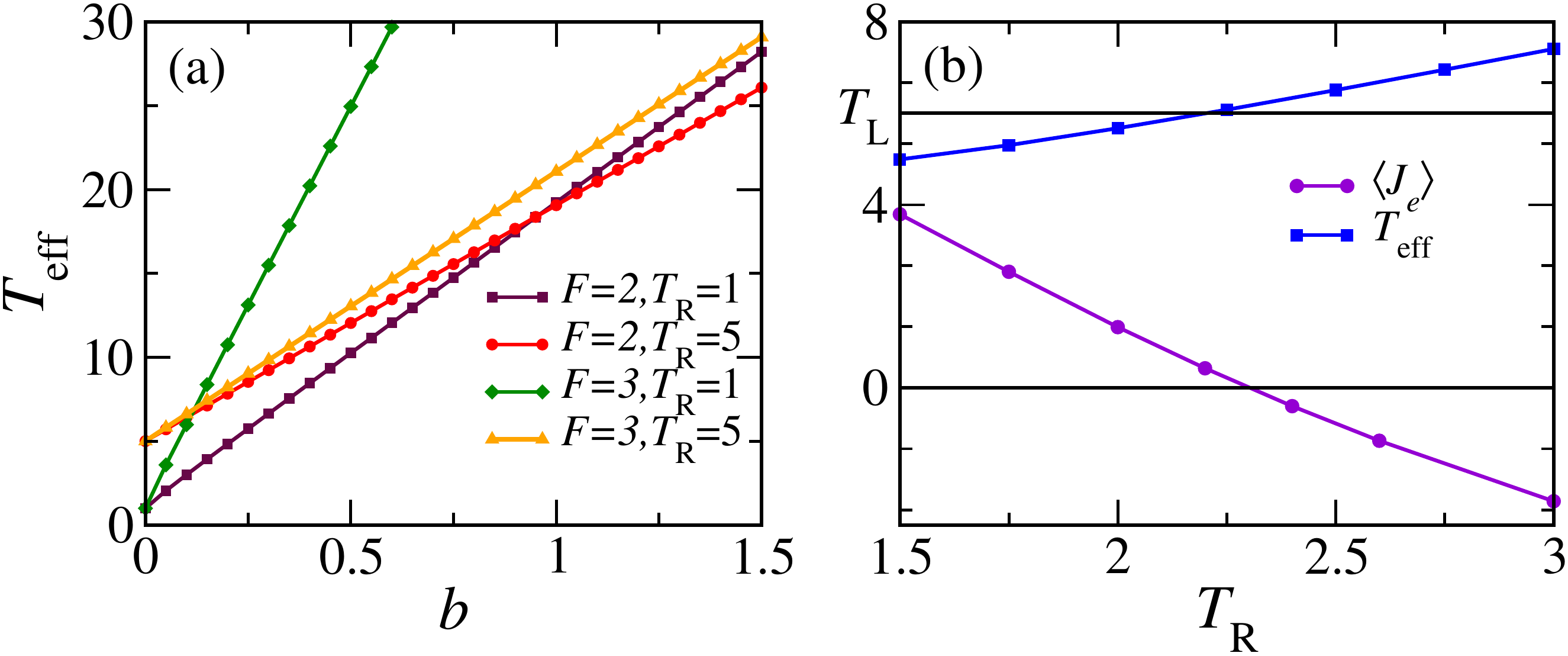} 
\caption{ (a) The effective temperature $T_\text{eff}$ defined in \eqref{eft} for a three-level system, increasing with equalizing strength $b$. The case $b=0$ corresponds to  equilibrium where $T_\text{eff} = T$. The crossing of two curves at the same $F$  but with different $T_R$ is a signature of the resulting population inversion. (b) Illustrating the meaning of effective temperature: when the right reservoir has an effective temperature that equals the temperature of the left reservoir, the energy current gets close to zero and then reverses sign. Here $b=0.5, F=1$ and the left temperature $T_L=6.$ The current $\la J_e \ra$ has been amplified by a factor for better visibility.
}
\label{fig:inv}
\end{figure}

This system can be thought of as a simple discrete model for energy transport in an extended system; the particles represent energy packets which are transported across the system without scattering. The situation is somewhat similar to the phenomenon of stimulated Raman adiabatic passage \cite{stirap}.
When the temperatures $T_L > T_R$ are different, there will be an energy current flowing through the system which is proportional to the temperature difference (Fourier law).\\

Here we are interested in the effect of population inversion on the thermal current. To this end we choose the jump rates at, say, the right boundary in a way so that the highest energy level $K$ becomes most populated at low temperature $T_R.$ We also assume free particle motion at the boundaries. The complete dynamics is then characterized by the transition rates $k(i,\ell; j,m )$ of a single particle from position $(i,\ell)$ to position $(j,m):$
\bea
\text{Bulk} ~~~ \quad \quad \quad \quad \quad \;\; \; k(i,\ell; i\pm 1,\ell ) &=& 1 \quad \forall i \ne 1,L \cr
\text{Left Boundary} ~~~ \quad \quad \quad k(1,\ell; 2,\ell ) &=& n_{1\ell} \cr
k(1,\ell; 1,\ell+1 ) &=& n_{1\ell} ~e^{-\frac 1{T_L} (E_{\ell+1} -E_\ell)}  \quad \forall \ell \ne K \cr
k(1,\ell; 1,\ell-1 ) &=& n_{1\ell} \quad  \quad\quad  \quad \quad  \quad \quad  \; \forall \ell \ne 1 \cr 
\text{Right Boundary} ~~~ k(L,\ell; L-1,\ell) &=& n_{L\ell} \cr 
k(L,\ell; L,\ell+1 ) &=& n_{L\ell}~ e^{-\frac 1{T_R} (E_{\ell+1} -E_\ell)} ~~ \forall \ell \ne K-1, K \cr
                          k(L,\ell; L,\ell-1 ) &=& n_{L\ell} \quad \qquad \quad \qquad \; \forall \ell \ne K \cr
k(L,K-1; L,K ) &=& n_{L,K-1}~ e^{-\frac F{T_R}} e^{-\frac 1{T_R} (E_{K} -E_{K-1})}  \cr
k(L,K; L,K-1 ) &=& n_{L,K}~ e^{-\frac F{T_R}}  \cr
k(L,1; L,K) &=& b~ n_{L,1}  \cr
k(L,K; L,1) &=& b~ n_{L,K} \label{eq:zrp_dyn}
\eea
The dynamics conserves the particle density $\rho$ in the system but not the total energy. \\
\\
The equalizing symmetric prefactor $b>0$ between the highest and lowest energy level along with the factor $e^{-F/T_R}$ at the right boundary gives rise to the desired population inversion \cite{Winny}. 
To illustrate that effect we first concentrate on the right reservoir itself --- decoupled from the bulk. It is equivalent to a system with $K$ energy levels where the lowest and highest levels are connected by an equalizer and an additional energy barrier $F$ exists between the $K$ and $K-1^{th}$ level. A good indicator of the resulting population inversion is the effective temperature measured from the stationary state probability of the highest and lowest levels, here defined as
\bea\label{eft}
T_\text{eff} = (E_K -E_1) \left[\log \frac{\rho_1}{\rho_K}\right]^{-1}
\eea
Fig. \ref{fig:inv} shows that effective temperature $T_\text{eff}$, increasing with the strength of the equalizer $b$. For a fixed $F,$ the curve of $T_\text{eff}$ corresponding to a lower thermodynamic temperature $T_R$ crosses that of a higher temperature from below signifying population inversion. The barrier $F$ facilitates this phenomenon --- crossing occurs for smaller $b$ when $F$ is increased.

\begin{figure}[th]
 \centering
 \includegraphics[width=14 cm]{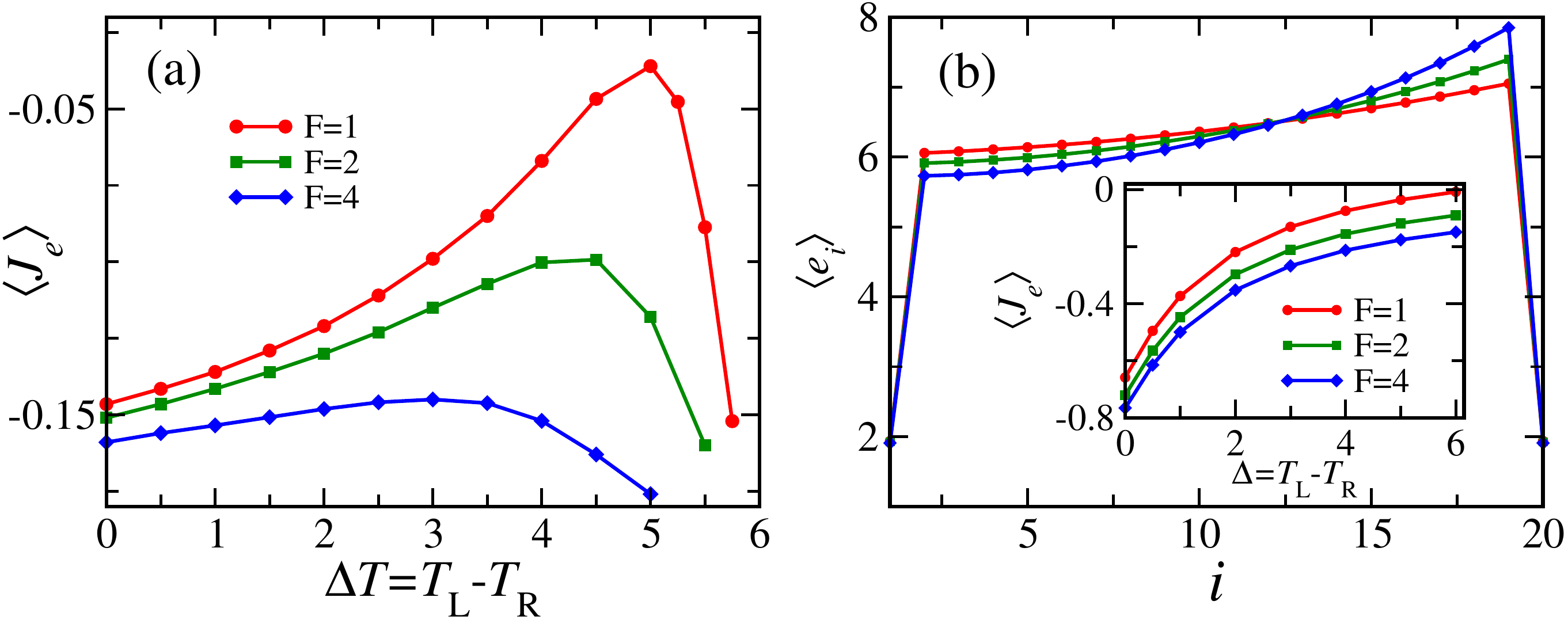}
 % mult_zrp.pdf: 595x842 pixel, 72dpi, 20.99x29.70 cm, bb=0 0 595 842
 \caption{(a) Energy current $\la J_e \ra$ as a function of the temperature difference $\Delta T=T_L-T_R$ for different values of $F.$ Here the temperature gradient is increased by decreasing $T_R$ while $T_L=6$ is kept fixed. (b) The energy profile $\la e_i \ra$ for different values of $F$  for fixed $T_L=6, T_R=1.$ 
 The inset shows $\la J_e \ra$ as a function of the temperature difference $\Delta T=T_L-T_R$ when $T_R=1$ is fixed.  Here $b=1$ and the number of energy levels $K=3$ and particle density $\rho=2.0.$ Also, we have taken $E_\ell = (l-1).$ }
 \label{fig:cur}
\end{figure}

\noindent One can expect that when this system serves as a thermal reservoir it would drive a current opposite to the thermal gradient which is present  even when $T_L=T_R.$ 

%The equalizer introduces an additional nonequilibrium in the system which drives a current opposite to the thermal gradient which is present  even when $T_L=T_R.$ 
\noindent The quantity of interest is the total energy current through the system,
\bea
J_e = \sum_{i=1}^{L-1} \sum_{\ell=1}^K E_\ell J^{i,i+1}_\ell
\eea
where $J^{i,i+1}_\ell$ denotes the particle current (net number of jumps) from $i \to i+1$ at the $\ell^{th}$ level.  We have adopted the notation where the superscript to the current $J$ indicates the spatial position and the subscript indicates the energy level. Fig. \ref{fig:inv}(b) illustrates  the relevance of the effective temperature for the full system. The current flowing through the system vanishes when  $T_\text{eff} \simeq T_L$ $i.e$ when the effective bias becomes zero. \\

\noindent To calculate the response we need the usual path dependent entropy and dynamical activity,
\bea
S_{T_R}(\omega) &=& - \frac 1{T_R} \sum_{\ell=1}^{K-1} J^R_{\ell,\ell+1} \Delta E_\ell \cr
D_{T_R}(\omega) &=& \frac 1{2T_R} \left[\sum_{\ell=1}^{K-1} I_{\ell,\ell+1} \Delta E_\ell  + 2F I_{K-1,K} \right] \cr
&+& \int_0^t \id \tau \left[  \sum_{\ell=1}^{K-2} e^{\left[- \frac{\Delta E_\ell}{T_R}\right]} n_{L,l}(\tau) +  e^{\left[- \frac{\Delta E_{K-1}+F}{T_R}\right]} n_{L,K-1}(\tau) +  e^{\left[- \frac{F}{T_R}\right]} n_{L,K}(\tau) \right] \n
\eea
Here $J^R_{\ell,\ell+1}$ and $I_{\ell,\ell+1}$ are the net particle current (directed from $\ell$ to $\ell+1^{th}$ level) and traffic (total number of jumps)  respectively at the right boundary between $\ell^{th}$ energy level and $\ell+1^{th}$ level; $\Delta E_\ell = E_{\ell+1}-E_\ell$ is the corresponding energy difference between these levels. Note that since we are interested in the thermal response where the thermal gradient is changed by varying  the right bath temperature $T_R$ we have kept only those terms which depend on $T_R.$  The thermal conductivity can now be calculated from formula \eqref{eq:dQSD_cov} where $S'$ and $D'$ would be the derivatives with respect to $\Delta= T_L-T_R.$ Once more we see that the quantity $F$ which is related to the reactivity of a particular transition appears explicitly only in the dynamical activity. \\

It is easy to see that this dynamics is essentially an inhomogeneous zero range process \cite{zrp} on a two--dimensional lattice where the hop rate   $u_{\alpha \gamma}(n_\alpha)$ from site $\alpha$ to $\gamma$ depends only on the particle number at the departure site. It is known that the stationary distribution has a product structure when the hop rate has the general form
\bea
u_{\alpha \gamma}(n_\alpha) = u_\alpha(n_\alpha) W_{\alpha \gamma}
\eea
where $W_{\alpha \gamma}$ gives the probability that $\gamma$ is chosen as the target site; $\sum_\gamma W_{\alpha \gamma}=1.$
The dynamics \eqref{eq:zrp_dyn} satisfies this condition with $u_\alpha(n) = \xi_\alpha$ when $\alpha=\{i,\ell\}$ refers to a site in the bulk and $u_\alpha(n) = n \xi_\alpha$ for boundary sites; $\xi_\alpha$ being the $n$--independent part of the escape rate of site $\alpha$, containing the energy dependent component of the transition rates. For example $\xi_{i,\ell}=2$ for bulk sites, $\xi_{L,1} = 1+b+\exp{(-\Delta E_1/T_R)}$ and so on.  With this identification and following \cite{zrp} one can write the individual site weights,
\bea
f_\alpha(n) =  \prod_{m=1}^n \left[\frac{q_\alpha}{u_\alpha(m)}\right]
\eea
where $q_\alpha$ are the solutions of the linear system of equations $q_\alpha = \sum_\gamma W_{\gamma \alpha} q_\gamma.$ \\
%It is convenient to work in the grand-canonical picture where the fugacity $z$ fixes the total particle density $\rho$ of the system. 
The stationary current and density profile $\rho_{i\ell}$ (satisfying $\frac 1{LK}\sum_{i,\ell} \rho_{i\ell} = \rho $) can easily be obtained from this exact solution. The average spatial energy profile, defined as
\bea
\la e_i \ra = \sum_{\ell=1}^K E_\ell \rho_{i\ell}
\eea
is plotted in Fig. \ref{fig:cur}(b). 

The average energy current $\la J_e \ra$ as a function of the temperature gradient $\Delta$ for different values of $F$ is shown in Fig. \ref{fig:cur}(a). While the thermal gradient drives a current from the left boundary to the right one, because of the population inversion, there is an opposing current due to the density gradient at the highest level. Now, the population inversion becomes more efficient as the right temperature $T_R$ goes down, hence if the thermal gradient is increased by decreasing $T_R$ the opposing drive becomes stronger and hence the current goes down.   
%This is seen in Fig. \ref{fig:cur}(a) where the energy current is plotted as a function of the temperature difference. 
%This is what we see in the figure where the current goes down when the temperature difference becomes large. On the 
On the other hand we also expect from the above explanation that there is no negative response when the left temperature $T_L$ increases. That is confirmed and shown in the inset of Fig. \ref{fig:cur}(b).

\section{Conclusion}
Non-thermodynamic features appear naturally under nonequilibrium conditions.  We have given examples how time-symmetric kinetic aspects contribute to interesting phenomenology.  The new unifying quantity here is the dynamical activity which via its excess due to the perturbation correlates from second order onwards with the (excess) entropy flux and produces the frenetic contribution to response.  Frenetic aspects arise then for example from reactivities that are modified by external fields or by the nonequilibrium condition in general.

%%%%%%%%%%%%%%%%%%%
%%%%%%%%%%%%%%%%%%%%%%%%%%%%%%%%%%%%%%%%%%%%%%%%%%%%%%%%%%%%%%%%%%%%%%
\ack
%%%%%%%%%%%%%%%%%%%%%%%%%%%%%%%%%%%%%%%%%%%%%%%%%%%%%%%%%%%%%%%%%%%%%% 
We very much thank the organizers for a great conference and stimulating environment.
%%%%%%%%%%%%%%%%%%%%%%%%%%%%%%%%%%%%%%%%%%%%

\end{document}